# Plasmonic nanoprism distributions to promote enhanced and uniform energy deposition in passive and active targets

Dávid Vass[1,2], Emese Tóth[1,2], András Szenes[1,2], Balázs Bánhelyi[2,3], István Papp[2,4], Tamás Biró[2], László Pál Csernai[2,4,5], Norbert Kroó[2,6], and Mária Csete[1,2,*]

[1] *University of Szeged, Department of Optics and Quantum Electronics, Szeged 6720, Hungary*
[2] *Wigner Research Centre for Physics, Budapest 1121, Hungary*
[3] *University of Szeged, Department of Computational Optimization, Szeged 6720, Hungary*
[4] *University of Bergen, Department of Physics and Technology, Bergen 5007, Norway*
[5] *Frankfurt Institute for Advanced Studies, Frankfurt/Main 60438, Germany*
[6] *Hungarian Academy of Sciences, Budapest 1051, Hungary*
*\*mcsete@physx.u-szeged.hu*

**Abstract**

Passive and active targets, implanted with gold nanoprisms, were designed to achieve enhanced and uniform power absorption during two-sided illumination by short laser pulses. The target length was adjusted to match the short laser pulse-length. Capabilities of three different, uniform, single-peaked Gaussian and adjusted, nanoresonator number density distributions were compared. The average local **E**-field inside the gain medium and on the surface of the nanoprisms were mapped as a function of the pump **E**-field strength and dye concentration, assuming a uniform nanoresonator distribution. The optimal parameters were adopted to each inspected nanoprism distributions. The time-evolution of the near-field enhancement (*NFE*), integrated power-loss and deposited energy were determined, additionally, the time-evolution of the standard deviation of these quantities was monitored. A comparative study was performed on passive and active targets, to determine the most advantageous nanoprism number density distribution type and to consider the advantages of dye doping. Based on the results, the adjusted distribution is proposed both in passive and active targets. Doping with the dye is advantageous in every inspected distribution in decreasing the minimal standard deviation of the *NFE*. It is advantageous in decreasing the delay of the minimal standard deviation in the power-loss and deposited energy, the standard deviation of the *NFE* as well as in increasing the FOM of the *NFE* in the uniform and adjusted distributions. In addition, doping allows for decreasing the delay of the minimal standard deviation in the *NFE* / increasing the mean *NFE* / decreasing the standard deviation of the power-loss and deposited energy in the uniform / Gaussian / adjusted distribution.

**Introduction**

The control of the nuclear reactions could address the energy demand, however, sustainable technology to achieve stable fusion has not been developed yet [1]. Various approaches in nanophotonics are focused on enhancing charged particles' density [2], accelerating them [3,4], extending the cut-off energy of crucial phenomena [5,6] and improving the conversion efficiency [7]. Most of these approaches rely on boosting the near-field, through better confinement and enhancement attainable via various nanophotonic resonators.

Large and confined near-field can be achieved through plasmonic nanoparticles (NPs) of different types, due to the localized surface plasmon resonance (LSPR). The **E**-field enhancement achievable via plasmonic nanoantennas and their patterns can reach several orders of magnitude [8,9]. The degree and distribution of the **E**-field enhancement originating from LSPR can be tuned by varying the parameters of plasmonic nanoresonators, including the shape, size, composition, as well as the embedding environment [10]. LSPRs can be excited on asymmetric nanoantennas, e.g. nanoprisms, which offer the specific advantage of extreme **E**-field localization in a single hot-spot [11].

In the case of individual triangular nanoantennas, the efficient excitation of LSPR requires **E**-field oscillation along their long axis, allowing for the strongest **E**-field confinement at the termination with the smallest radius of curvature [12].

An almost uniform size distribution of nanoprisms can be achieved through chemical procedures or combined laser and colloid-sphere lithographies [11]. Moreover, ordered patterns of uniform and oriented nanoprisms show a promise of enhancing the **E**-field via surface lattice resonances (SLR) that can be excited, when the period is comparable to the wavelength in the specific embedding medium. Such patterns of nanoprisms with controlled location and orientation can be fabricated using nanosphere lithography [13].

Plasmonic nanoparticles (NP) can be used in wide fields of multidisciplinary applications. As drug delivery systems, these NPs offer competitive methods for cancer treatment by increasing the therapies' effectiveness and facilitating the overcoming of various challenges, such as drug resistance [14]. They are also applicable in nanoparticle-based bone tissue engineering, by contributing to increased bone regeneration efficacy [15]. Nanoparticles can serve as anatomic and molecular imaging markers as well, due to their small size and high surface-to-volume ratio. Moreover, they offer stable and intense imaging signals, with multimodal and multiplexing capabilities [16].

Due to the large and confined local **E**-fields, nanoparticles can enhance the spontaneous emission of nearby emitters [17]. Accordingly, plasmonic nanoparticles can be used to increase the luminescence of dye molecules, moreover plasmon enhanced lasing can be also achieved [18,19]. Quantum dots (QDs) can also effectively couple with metal nanoprisms, in case of proper geometry tuning. When the surface plasmon resonance wavelength overlaps with the photoluminescence spectrum of the QDs, the emission intensity can be increased, while the lifetime can be decreased according to the Purcell effect [20]. Plasmon enhanced emission phenomena offer a tool to further boost the **E**-field strength.

In addition to individual nanoresonators, various multitudes of nanoparticles were studied. Random lasing can be achieved with randomly positioned resonant scatterers inside an optical gain medium. In these systems the lasing properties are determined by considering the interplay between the gain medium and the scattering centers [21]. Random lasing action can be enhanced with metal nanoparticles, as the scattering cross-section is increased compared to the geometrical cross-section due to the surface plasmon resonance, while the gain volume is decreased due to the **E**-field confinement. The large localized **E**-field in nanoresonator integrated media ensures control over both absorption and emission phenomena, leading to a considerable fluorescence enhancement [22]. In gain media seeded with nanoprisms, multiple emission spikes appear, and the lasing threshold decreases compared to media without nanoresonators. Pronounced full-width-at-half maximum (*FWHM*) narrowing can be achieved in the presence of oriented nanoparticles due to the coherent feedback, and the threshold is significantly decreased due to the increased local **E**-field and corresponding local pump fluence [23]. Experimental evidence of random lasing action was demonstrated using nanoprisms embedded into a substrate. The emission wavelength can blue-shift due to the strain stemming from the bending of the polymer target, allowing for tunable lasing emission [24].

Plasmonic nanoparticles can be used for various light controlled energy deposition purposes [25.27]. In our previous study it was shown that by optimizing the distribution of the NPs along an extended target, uniform energy deposition can be achieved, which is crucial in fusion applications [28].

The possibility of balancing the deposited energy along an extended target was demonstrated using core-shells and nanorods, both in passive and active media [28,29]. When extended targets are illuminated by short laser pulses, the time-evolution of the induced phenomena is governed by the laser pulse shape.

It was demonstrated that to achieve uniformly distributed near-field enhancement, double-sided illumination is advantageous [30]. This is an important tool to avoid instabilities that can arise in the case of three-fold, namely spectral-spatial-temporal, laser pulse confinement.

In this study a theoretical investigation is performed based on the idea that the plasmonic nanoprisms can be advantageous in achieving larger near-field enhancement and power-loss along the target. By modifying the nanoparticle number density distribution, uniform power absorption, near-field enhancement and energy deposition can be achieved.

**Methods**

Steady-state and time-domain computations were realized using the RF module of COMSOL Multiphysics. The target was a 21 μm thick polymer slab, divided into seven, uniformly 3 μm thick, consecutive layers. The target was seeded with 70 nanoprisms made of gold in random orientation and position (inset in Fig. 1). The target was illuminated by two counter-propagating 120 fs short-pulses, with a central wavelength of 795 nm, in order to make the results comparable with experiments that are in progress with Ti:Sapphire laser. The geometry of the nanoprisms was tuned to ensure resonance matching with the central wavelength of the laser pulse. The thickness of the nanoprisms was a predefined 10 nm, while the base of the triangular antenna was tuned to 82 nm in case of gold to ensure resonance at 795 nm. Three different nanoresonator distributions were examined: uniform, single-peaked Gaussian and adjusted distribution.

Beside the passive targets, their active counterparts were also considered, where the polymer was doped with a laser dye (LDS 798). A numerical pump-and-probe simulation was performed using a 532 nm monochromatic CW pump beam and a 795 nm CW probe beam, following the method detailed in our previous studies [19,29]. Via steady-state modelling the average local **E**-field on the surface of the nanoprisms and in the volume of the gain medium along the target was mapped over the pump **E**-field strength ($E_{pump}$) and dye concentration ($c$) parameter plane. Based on the maps taken primarily using a target with uniform nanoprism number density and dye molecule concentration distribution, a pump **E**-field strength of $2\times10^6$ V/m and a dye concentration of $3.25\times10^{26}$ m$^{-3}$ were used in the active target. With these pump and dye parameters, the local **E**-field is simultaneously enhanced both in the gain medium and on the surface of the nanoprisms (Fig. 1 a, d). It was proven that the local **E**-field was efficiently enhanced at the selected pump **E**-field strength and dye concentration, when the triangular nanoresonator distribution was modified from uniform to either Gaussian (Fig. 1 b, e) or to adjusted (Fig. 1 c, f) distribution. Importantly, the pump and probe intensity is slightly and considerably below the damage threshold of the nanoprisms, when the pump and probe **E**-field is $2\times10^6$ V/m and $10^4$ V/m, respectively. In case of the adjusted nanoprism distribution the dye molecule concentration distribution was also modified with the criterion that the average concentration of the dye remains the same as in the case of uniformly doped targets.

The time-evolution of the power-loss ($PL(t)$), deposited energy ($E(t)$) (see the Supplementary material) and average near-field enhancement ($NFE(t)$) was determined in each layer. The time-evolution of the standard deviation was inspected to determine the value ($\delta_{min\_PL/E/NFE}$), time-instant ($t_{min\_PL/E/NFE}$) and delay compared to the time-instant ($t_{overlap}$ = 240 fs) of theoretical overlap of counter-propagating pulses ($\Delta t_{min\_PL/E/NFE}$ = | $t_{overlap}$ - $t_{min\_PL/E/NFE}$ |) of the minimal standard deviation.

The power-loss (and also the deposited energy) was integrated until $t_{overlap}$. Based on the time-evolution of the integrated power-loss (as well as of the deposited energy) and *NFE,* the average values of these quantities were calculated at $t_{overlap}$, and the normalized standard deviation ($\delta_{PL}$, $\delta_{E}$, $\delta_{NFE}$) along the target was determined as follows:

$$\delta = \frac{standard\ deviation}{average\ value} \qquad (1)$$

The figure of merit (FOM) was defined as the ratio of the average value of the inspected quantity and the normalized standard deviation (*FOM*$_{PL}$ = PL/$\delta_{PL}$, *FOM*$_E$ = E/$\delta_E$, *FOM*$_{NFE}$ = NFE/$\delta_{NFE}$) (Supplementary Table S1). About all of the deposited energy-related data please see Supplementary material.

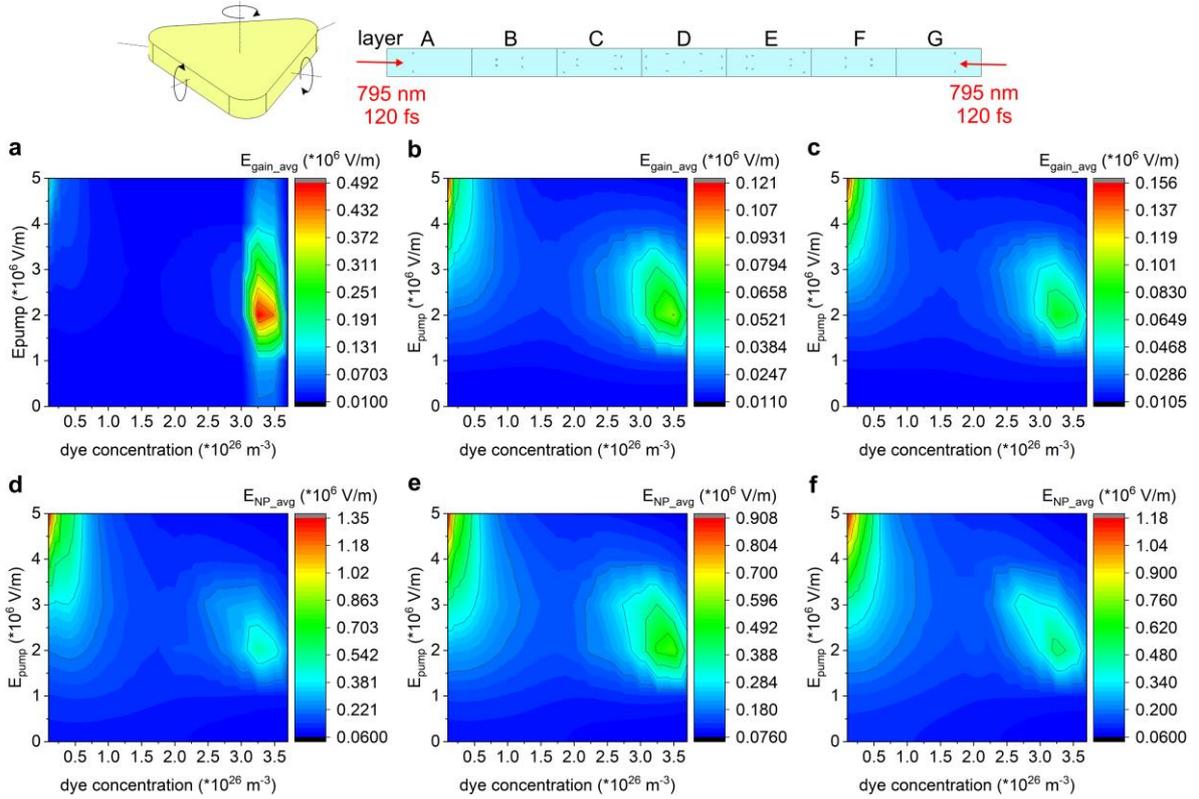

**Figure 1.** Schematic and the NFE maps of the inspected nanoprism distributions. The average local **E**-field (a-c) inside the volume of the gain medium and (d-f) on the surface of the nanoprisms as a function of the dye molecule concentration and the pump **E**-field strength ($E_{pump}$) in case of (a, d) uniform, (b, e) Gaussian and (c, f) adjusted nanoresonator distribution. Inset: a schematic figure of an individual nanoprism and the vertical cross-section of seven layers in a pulse-scaled target, implanted with nanoprisms and doped with dye optionally, illuminated by two horizontally counter-propagating short-pulses.

**Results and discussion**

**Passive targets**

*Dynamics of standard deviation*

In the passive targets the time-evolution of the power-loss and *NFE* in each layer is different (Supplementary Table S1, Fig. 2 and 3). The uniform number density distribution shows the largest minimal standard deviation in the power-loss with the largest delay, and intermediate minimal standard deviation in the *NFE* also with the largest delay (Fig. 2a, d, g and Fig. 3a, d, g). The Gaussian distribution is advantageous due to the intermediate minimal standard deviation with intermediate delay in the power-loss, but has the largest minimal standard deviation in the *NFE*, though with the smallest delay (Fig. 2b, e, h and Fig. 3b, e, h). The advantage of the adjusted distribution is that the smallest minimal standard deviation is achieved both in the power-loss and in the *NFE*, with the smallest and with compromised intermediate delay, respectively (Fig. 2c, f, i and Fig. 3c, f, i).

*Evaluation at the time-instant of the pulses'-overlap*

Intermediate averaged integrated power-loss is achieved with the uniform nanoresonator distribution, but the standard deviation is the largest at the time-instant of counter-propagating pulses' overlap (Supplementary Table S1, Fig. 2. g, j). In the average *NFE* the uniform distribution is weak, as it allows for the smallest average value (3.94-fold enhancement) and the largest standard deviation (Supplementary Table S1, Fig. 3. g, j). In case of the Gaussian distribution the smallest power-loss with an intermediate standard deviation can be achieved, while the *NFE* (6.55-fold) and its standard deviation is intermediate (Supplementary Table S1, Fig. 2. h, k and Fig. 3. h, k). The adjusted distribution produces the largest power-loss, and *NFE (8.61-fold)*, with the smallest standard deviation in both quantities (Supplementary Table S1, Fig. 2. i, l and Fig. 3. i, l).

Based on the FOM of the power-loss, the uniform distribution ($3.58 \times 10^{-17}$ J) is intermediate, the least advantageous is the single-peaked Gaussian distribution ($1.68 \times 10^{-17}$ J), while the adjusted distribution ($7.14 \times 10^{-16}$ J) is the most advantageous. However, comparing the FOM of the *NFE* the relations modify the ranking of the targets. The uniform distribution (8.54) becomes the weakest, the Gaussian distribution (18.29) is intermediate, while the adjusted distribution (26.08) remains the most advantageous also the in the *FOM$_{NFE}$* (Supplementary Table S1).

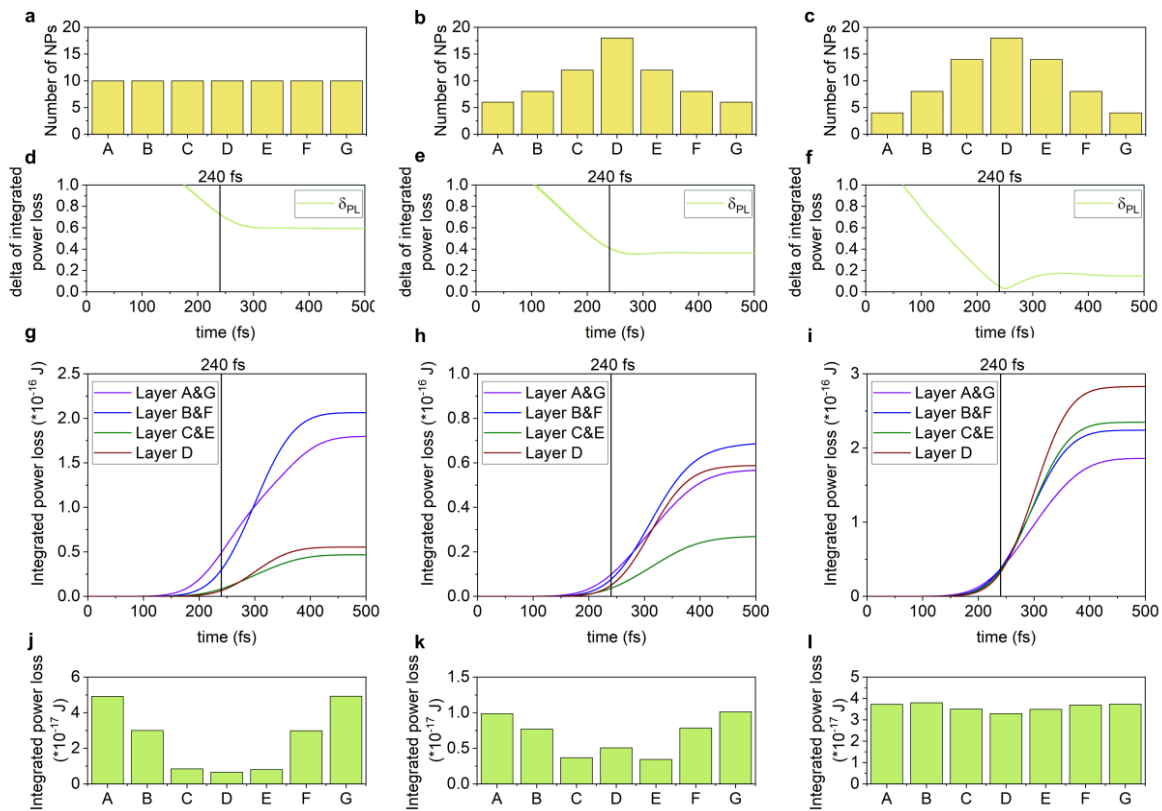

**Figure 2.** Time-dependent power-loss in passive targets. The (a-c) nanoprism number density distribution along the target, the time-evolution of the (d-f) standard deviation of the power-loss and the (g-i) integrated power-loss. (j-l) The distribution of the integrated power-loss in different layers at 240 fs. (a, d, g, j) uniform, (b, e, h, k) Gaussian and (c, f, i, l) adjusted distributions.

The uniform distribution is intermediate in the power-loss and in the FOM of the power-loss, while it is the weakest in the average *NFE,* in the standard deviation of the power-loss and *NFE* as well as in the FOM of the *NFE*.

The Gaussian distribution is intermediate in the standard deviation of the power-loss and *NFE* at 240 fs, in the average *NFE* and in the FOM of the *NFE*, while it is the weakest in the achieved power-loss and in the FOM of the power-loss.

The adjusted distribution achieves the smallest standard deviation, the largest average value and also the largest FOM of the power-loss and *NFE*.

*Ranking of the passive targets*

If every inspected quantity is equally considered in the ranking, namely, all quantities are counted, then the distributions are not comparable. The weakest / compromised intermediate / the most preferable distribution is the uniform / Gaussian / adjusted, as it shows 7 – 3 – 0 / 3 – 6 – 1 / 0 – 1 – 9 / quantities, in which the specific system is the weakest – intermediate – the most preferable. Based on the power-loss / *NFE* FOM, the distribution ranking shows Gaussian / uniform – uniform / Gaussian – adjusted / adjusted order, so the most advantageous is the adjusted distribution, in accordance with the intuitive expectations (Supplementary Table S2 and S3). The non-uniform distributions possess better characteristic already in passive targets.

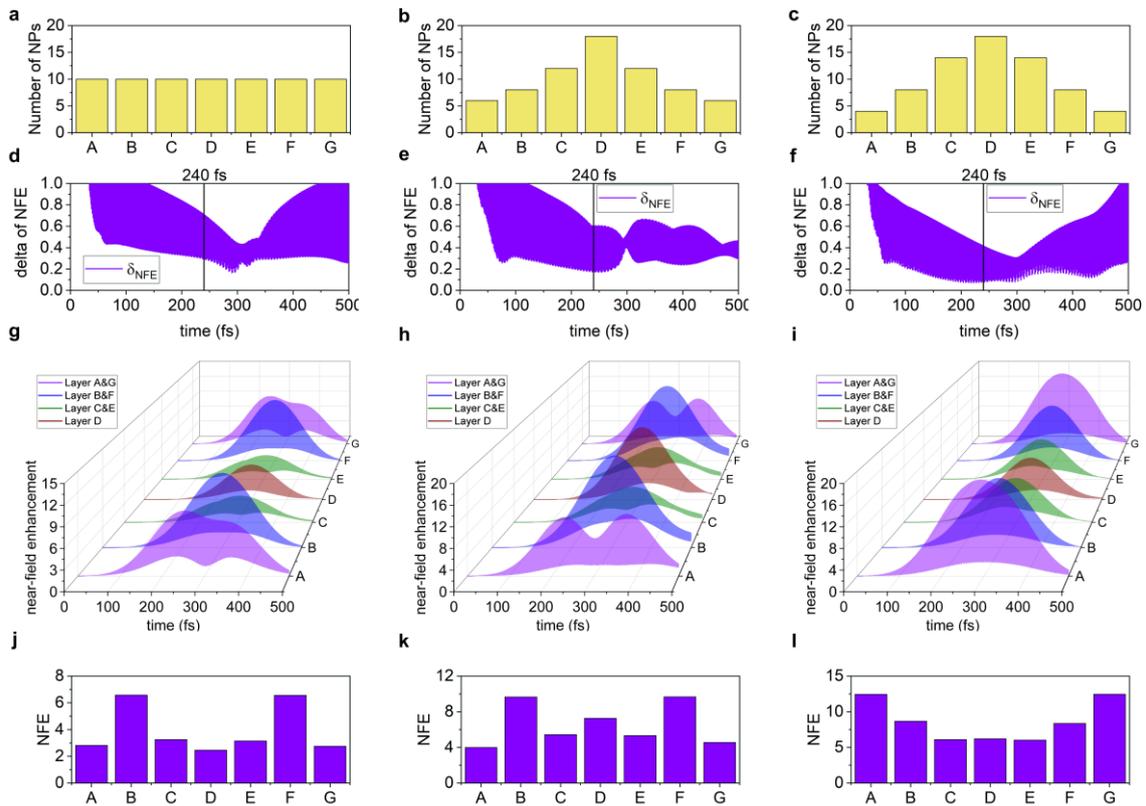

**Figure 3.** Time-dependent near-field enhancement in passive targets. The (a-c) triangular nanoresonator number density distribution along the target. The time-evolution of the (d-f) standard deviation of the *NFE* and the (g-i) instantaneous *NFE*. (j-l) The distribution of the *NFE* in different layers at 240 fs. (a, d, g, j) uniform, (b, e, h, k) Gaussian and (c, f, i, l) adjusted distributions.

**Active targets**

*Dynamics of standard deviation*

In the active targets the time-evolution of the power-loss and *NFE* in each layer is different as in the case of nanoprism distributions in passive targets (Supplementary Table S1, Fig. 4 and 5).

The uniform distribution in active target shows the largest minimal standard deviation in the power-loss with intermediate delay, and intermediate minimal standard deviation in the *NFE* with intermediate delay (Fig. 4a, d, g and Fig. 5a, d, g). Compared to the passive uniform distribution the minimal standard deviation is considerably increased in the power-loss, while it is slightly decreased in the *NFE*. The delay decreased both in minimal standard deviation in the power-loss and in the *NFE*, compared to the uniform nanoprism distribution in passive target (Supplementary Table S1).

In case of the single-peaked Gaussian distribution in active target, the minimal standard deviation of the power-loss is intermediate, but with the largest delay, while the minimal standard deviation of the *NFE* is the largest, though it is taken on advantageously with the smallest delay (Fig. 4b, e, h and Fig. 5 b, e, h). Compared to the Gaussian distribution in the passive target, the minimal standard deviation is considerably increased in the power-loss, and it is slightly decreased in the *NFE*, while the delay is considerably / slightly increased for the minimal standard deviation in the power-loss / *NFE* (Supplementary Table S1).

The advantage of the adjusted distribution in active target is that the smallest minimal standard deviation in the power-loss is achieved with the smallest delay, moreover the smallest standard deviation is reached in the *NFE*, though with the largest delay among the active targets. These results indicate that the temporal characteristics are predominantly improved in the adjusted distribution. Compared to the adjusted distribution in passive target, the minimal standard deviation is slightly increased in the power-loss and it is significantly decreased in the *NFE*. Although, only the latter is improved, both are the smallest among the minimal deviations taken in active targets. The delay of the minimal standard deviation is decreased in the power-loss compared to the adjusted distribution in passive target, while it is significantly increased in the *NFE*. The former indicates that the dye doping might be advantageous in terms of delay as well (Supplementary Table S1).

Dye doping of targets embedding uniform and adjusted nanoprism distributions is advantageous in decreasing the delay between the time-instant of minimal standard deviation and the theoretical overlap of counter-propagating pulses in case of the power-loss and *NFE* (as well as of the deposited energy, see supplementary material), except the $\Delta t_{min\_NFE}$ in the adjusted distribution, as the delay of the minimal standard deviation in the *NFE* is increased compared to its counterpart registered in case of passive target. The single-peaked Gaussian distribution is remarkably different, as the dye doping results in increased delay of the minimal standard deviation of all quantities.

Dye doping is also advantageous considering the decreased minimal standard deviation of *NFE*, however improvement of the uniformity of the power-loss in active targets is the subject of further study (Supplementary Table S1 and S2).

*Evaluation at the time-instant of pulse-overlap*

Similarly to the passive target, intermediate average integrated power-loss is achieved by the uniform nanoprism distribution, though with the largest standard deviation, while the average *NFE* (3.23-fold) is the smallest, with intermediate standard deviation at the theoretical overlap of the pulses (Supplementary Table S1, Fig. 4. g, j and Fig. 5. g, j).

Compared to the uniform distribution in passive target, both the power-loss and the *NFE* are slightly decreased, the standard deviation of the power-loss is increased, while for the *NFE* it is decreased (Supplementary Table S1).

The smallest power-loss is observed in case of the Gaussian distribution, with intermediate standard deviation, while the largest *NFE* (6.83-fold) can be achieved, however it is compromised with the largest standard deviation (Supplementary Table S1, Fig. 4. h, k and Fig. 5. h, k).

Compared to the Gaussian distribution in the passive target, the power-loss is slightly decreased, though the *NFE* is slightly increased, the standard deviation of both quantities is increased (Supplementary Table S1).

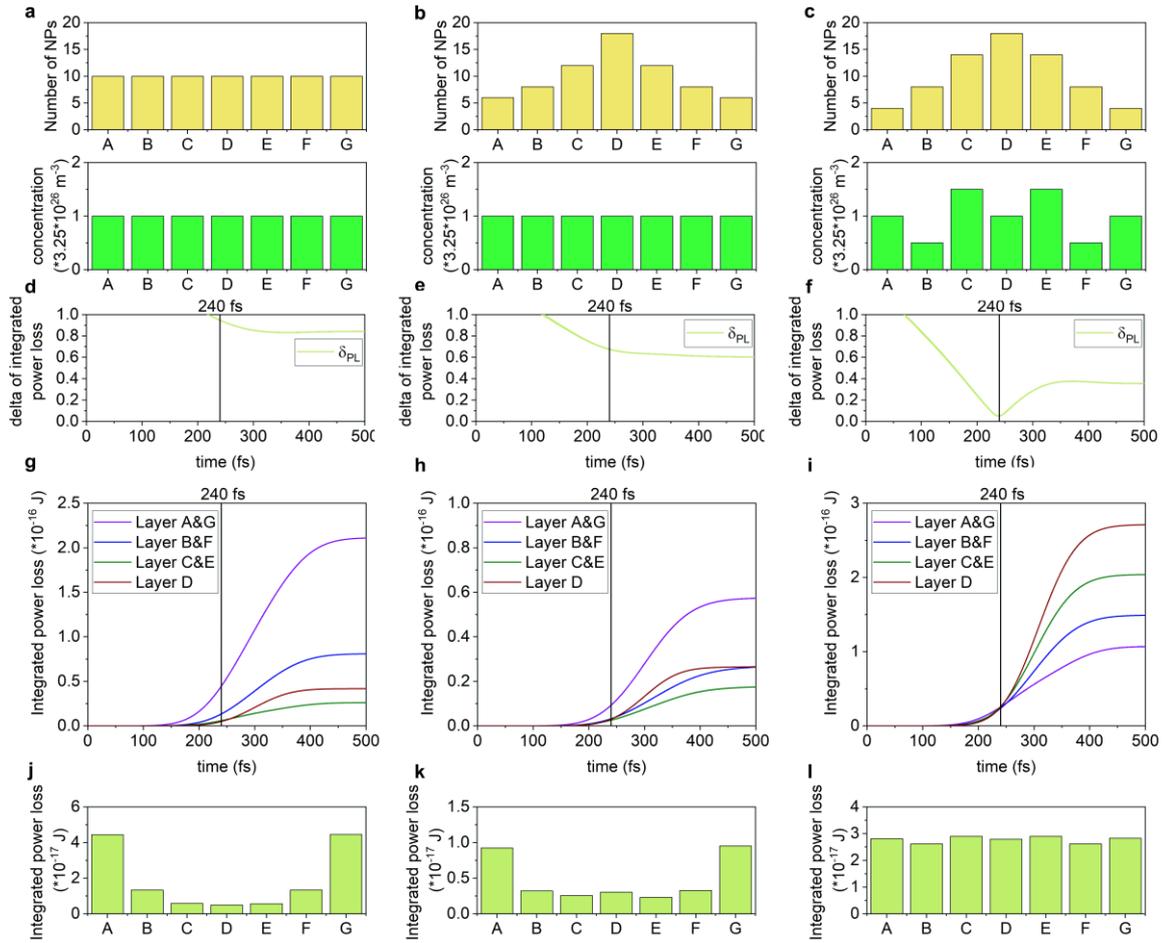

**Figure 4.** Time-dependent power-loss in active targets. The (a-c) nanoprism number density and dye molecule concentration distribution along the target. The time-evolution of the (d-f) standard deviation of the power-loss and the (g-i) integrated power-loss. (j-l) The distribution of the integrated power-loss in different layers at 240 fs. (a, d, g, j) uniform, (b, e, h, k) Gaussian and (c, f, i, l) adjusted distributions.

The adjusted distribution allowed for the largest power-loss with the smallest standard deviation, though it exhibited intermediate *NFE* (6-74-fold), but similarly with the smallest deviation (Supplementary Table S1, Fig. 4. i, l and Fig. 5. i, l). Compared to the adjusted distribution in passive target, the power-loss and *NFE* are slightly decreased, the standard deviation of the power-loss / *NFE* is slightly / significantly decreased, which indicates that the dye doping has well-defined advantages in both quantities in the achievement of larger degree uniformity along the target (Supplementary Table S1).

In case of uniform and Gaussian distributions, the integrated power-loss decreases, while the standard deviation increases compared to their counterparts in passive target. The power-loss also decreases in the adjusted target, but its standard deviation also decreases, the latter indicates that the dye seeding can be advantageous in the achievement of uniformity in the power-loss along the target (Supplementary Table S1 and S3).

Compared to their passive counterparts, the *NFE* and its standard deviation are decreased in the uniform and adjusted distributions, while in the Gaussian distribution both quantities are increased in active target (Supplementary Table S1 and S3).

These results indicate distribution dependent, compromised advantages of the dye doping, that manifest themselves in complementarily improvement either in the standard deviation or in the value of the *NFE*.

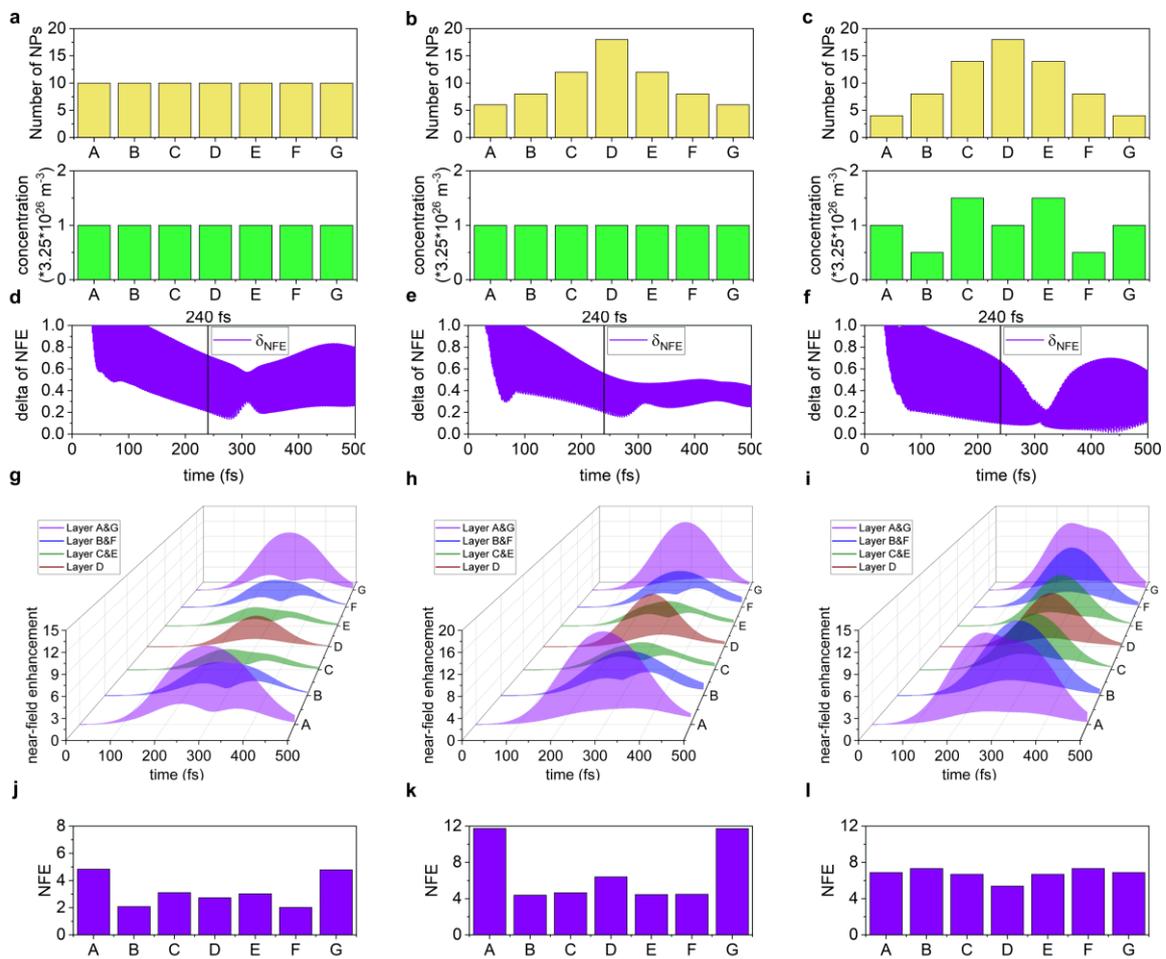

**Figure 5.** The time-dependent near-field enhancement in active targets. The (a-c) nanoprism number density and dye molecule concentration distribution along the target. The time-evolution of the (d-f) standard deviation of the *NFE* and the (g-i) instantaneous *NFE*. (j-l) The distribution of the *NFE* in different layers at 240 fs. (a, d, g, j) uniform, (b, e, h, k) Gaussian and (c, f, i, l) adjusted distributions.

Based on the FOM of the power-loss the uniform distribution is intermediate (1.99×10$^{-17}$ J), the Gaussian distribution is the weakest (7.02×10$^{-18}$ J), while the adjusted distribution is the most advantageous (5.76×10$^{-16}$ J). This ranking modifies, when the FOM of the *NFE* is compared, similarly to the passive targets. The uniform distribution becomes the least advantageous (8.97), the Gaussian distribution is intermediate (13.63), while the adjusted distribution remains the most beneficial (72.35) (Supplementary Table S1).

Compared to their counterpart distributions in the passive target, every FOM is decreased, except the FOM of the *NFE* in the uniform and adjusted distributions, which indicates a few advantages of dye seeding. In the active targets the ranking of the different inspected distributions remains the same both in the FOM of the power-loss and *NFE,* as in the passive targets (Supplementary Table S1 and S3).

At 240 fs the uniform distribution is intermediate in the power-loss, FOM of the power-loss and the standard deviation of the *NFE*, while it is the weakest in the standard deviation of the power-loss, mean *NFE* and the FOM of the *NFE*.

The largest *NFE* is achieved by the Gaussian distribution among the active targets, while it is intermediate in the standard deviation of the power-loss and FOM of the *NFE*. However, the Gaussian distribution shows the smallest power-loss, FOM of the power-loss and the largest standard deviation in the *NFE*.

The adjusted distribution is the most advantageous due to the largest power-loss, smallest standard deviation in the power-loss and *NFE* and the largest FOM of the power-loss and *NFE*. It is intermediate only the average *NFE*.

Considering the power-loss, the passive targets outperform the active ones in the average value, the standard deviation and also the FOM, except the adjusted distribution in the active target, that possesses a standard deviation smaller than its counterpart in passive target. Considering the *NFE,* the dye doping is advantageous in increasing the average *NFE* in the Gaussian distribution, and it is also beneficial in achieving smaller standard deviation and larger FOM in the uniform and adjusted distributions compared to their counterpart distributions in passive targets.

*Ranking of the active targets*

Similarly to the passive targets the ranking is not balanced, when every inspected quantity is equally considered, the only difference is that the Gaussian / adjusted distribution becomes significantly / slightly less advantageous, than the uniform / Gaussian distribution. Namely, the weakest / compromised intermediate / the most preferable is the uniform / Gaussian / adjusted distribution, as it shows 4 – 6 – 0 / 5 – 3 – 2 / 1 – 1 – 8 quantities; in which the specific system is the weakest – intermediate – the most preferable. Based on the power-loss / *NFE* FOM, the distribution ranking shows Gaussian / uniform – uniform / Gaussian – adjusted / adjusted order, so the most advantageous is the adjusted distribution, in accordance with the intuitive expectations, analogously with passive targets (Supplementary Table S2 and S3). The non-uniform distributions possess better characteristic also in active targets.

**Conclusion**

A comparative study was realized on different nanoprism distributions embedded into passive and active targets. A specific parameter region was identified by sweeping the dye molecule concentration and pump **E**-field strength, where large average local **E**-field can be achieved both in the gain medium and on the nanoprisms' surfaces, in each inspected distribution.

Based on the analyzed quantities using uniform distribution in passive target is the least efficient method to ensure uniform power-loss and near-field distributions. However, single-peaked Gaussian distribution can be advantageous due to that the smallest delay of the minimal standard deviation in the *NFE* can be achieved.

The Gaussian distribution is also advantageous in the amount of the deposited energy. The adjusted distribution is the most advantageous, due to the smallest minimal standard deviation in all inspected quantities, the smallest delay of the minimal standard deviation in the power-loss and deposited energy, the largest integrated power-loss and *NFE* value and the smallest standard deviation at 240 fs in the power-loss, *NFE* and deposited energy. Furthermore, the largest FOMs can be achieved in every inspected quantity with the adjusted distribution in case of passive targets. Based on these results the adjusted nanoprism distribution is proposed in the passive targets (Supplementary Table S1-S3).

Similarly, in the active targets the uniform distribution is the least advantageous, though it becomes intermediate in more, and remains the weakest in less quantities, compared to the counterpart distribution in passive target.

The Gaussian distribution has several advantages, namely the smallest delay of the minimal standard deviation in the *NFE* and the largest *NFE* and deposited energy is achieved by using single-peaked Gaussian distribution. However, on overall the adjusted distribution is the most advantageous, similarly to the passive targets. This is due to that the adjusted distribution allowed for the smallest minimal standard deviation in all inspected quantities, the smallest delay of the minimal standard deviation in the power-loss and deposited energy, the largest integrated power-loss, the smallest standard deviation at 240 fs in all inspected quantities, and also the largest FOM of the power-loss, *NFE* and deposited energy. Based on these results the adjusted nanoprism distribution is proposed in the active targets as well (Supplementary Table S1-S3, about energy-related data please see supplementary material).

Doping with dye of the target embedding the nanoprism distribution is not uniformly advantageous compared to the passive counterparts. In case of the uniform and adjusted distributions the minimal standard deviation of the *NFE*, and the delay of the minimal standard deviation in the power-loss and deposited energy become smaller, moreover the standard deviation of the *NFE* at 240 fs is also smaller, while the FOM of the *NFE* is larger. In addition to this, in the uniform distribution the delay of the minimal standard deviation in the *NFE*, while in the adjusted distribution the standard deviation of the power-loss and deposited energy at 240 fs becomes smaller compared to their counterparts in passive target. In case of the Gaussian distribution, using dye is advantageous in facilitating smaller minimal standard deviation of the *NFE*, similarly to the other two distributions, as well as in allowing for larger mean *NFE* value at 240 fs exclusively (Supplementary Table S1-S3).

Comparing every inspected target types and distributions the passive target with adjusted distribution is proposed, when the target is seeded with asymmetric nanoprisms, which is closely followed by its active counterpart in the global ranking. The standard deviation of the power-loss and energy (*NFE*) at 240 fs is reduced as well as the $FOM_{NFE}$ and $FOM_E$ is increased in the uniform – single-peaked Gaussian - adjusted order as it is expected in both (passive) targets. The other characteristic values -including the minimal standard deviation of *NFE* (standard deviation of the *NFE* at 240 fs) as well as the $FOM_{PL}$ in (active) both targets – exhibit the single-peaked Gaussian – uniform - adjusted order. This can be explained by that the uniform and Gaussian distributions have more predefined distribution related constraints and thus only a compromised uniformity can be achieved. In case of the adjusted distribution the nanoparticle and dye distribution were adjusted to minimize the standard deviations measured at 240 fs, and thus to make the integrated power-loss and *NFE* as uniform and high as possible at the theoretical time of overlap. In the passive targets, the improved power-loss and *NFE* uniformity implies an increase both in integrated power-loss and in *NFE*, while in the active target only the power-loss uniformity improvement is accompanied by increased integrated power-loss.

This can be explained by that all active targets are compromised, with balanced advantages and disadvantages. By doping the targets with dye, the standard deviation of the *NFE* (power-loss and deposited energy) at 240 fs was reduced except the Gaussian distribution (in the adjusted distribution), but the power-loss, deposited energy and achieved *NFE* (except the Gaussian) was smaller, than in the counterpart distributions in passive target.

According to the composite objective function, the FOM was improved for all quantities compared to the uniform distribution in passive and active targets (except the *FOM$_{PL}$* in Gaussian distributions). The adjusted distribution in active targets outperform the uniform distribution in passive target in all quantities, except the delay of the minimal standard deviation in the *NFE*. Moreover, the adjusted distribution in active target outperforms even its counterpart in passive target in the minimal standard deviation of the *NFE*, in the delay of the minimal standard deviation in the power-loss and deposited energy, and in the standard deviation of all inspected quantities at 240 fs.

Moreover, slight / significant FOM$_{NFE}$ improvement is achieved for a specific distribution via dye doping in the uniform / adjusted distribution compared to their counterparts passive target.

Joint optimization with composite objective functions and adding more constraints is a subject of further studies to precisely tune the distributions in order to achieve specific criteria of applications.

**Data availability**

The datasets used and / or analyzed during the current study are available from the corresponding author on request.

**Acknowledgements**

This work was supported by the National Research, Development and Innovation Office (NKFIH) of Hungary, projects: "Optimized nanoplasmonics" (K116362), "Nanoplasmonic Laser Inertial Fusion Research Laboratory" (NKFIH-2022-2.1.1-NL-2022-00002) and "National Laboratory for Cooperative Technologies" (NKFIH-2022-2.1.1-NL-2022-00012) in the framework of the Hungarian National Laboratory program.

**Author Contributions**

D.V.: Methodology, validation, formal analysis, investigation, data curation, visualization, writing—original draft preparation. E.T.: Methodology, validation, formal analysis, investigation, data curation. A.Sz.: Methodology, formal analysis, visualization, writing. B.B.: Software. I.P.: Review and editing. T.B.: Supervision. L.P.Cs.: Supervision. N.K.: Supervision. M.Cs.: Conceptualization, supervision, investigation, writing—review and editing.

All authors have read and agreed to the published version of the manuscript.

**Additional Information**

**Competing interests:** The authors declare no competing interests.

# Supporting Information

Dávid Vass[1,2], Emese Tóth[1,2], András Szenes[1,2], Balázs Bánhelyi[2,3], István Papp[2,4], Tamás Biró[2], László Pál Csernai[2,4,5], Norbert Kroó[2,6], and Mária Csete[1,2,*]

[1] *University of Szeged, Department of Optics and Quantum Electronics, Szeged 6720, Hungary*
[2] *Wigner Research Centre for Physics, Budapest 1121, Hungary*
[3] *University of Szeged, Department of Computational Optimization, Szeged 6720, Hungary*
[4] *University of Bergen, Department of Physics and Technology, Bergen 5007, Norway*
[5] *Frankfurt Institute for Advanced Studies, Frankfurt/Main 60438, Germany*
[6] *Hungarian Academy of Sciences, Budapest 1051, Hungary*

**Energy deposition and silver distributions**

**Passive targets**

*Dynamics of standard deviation*

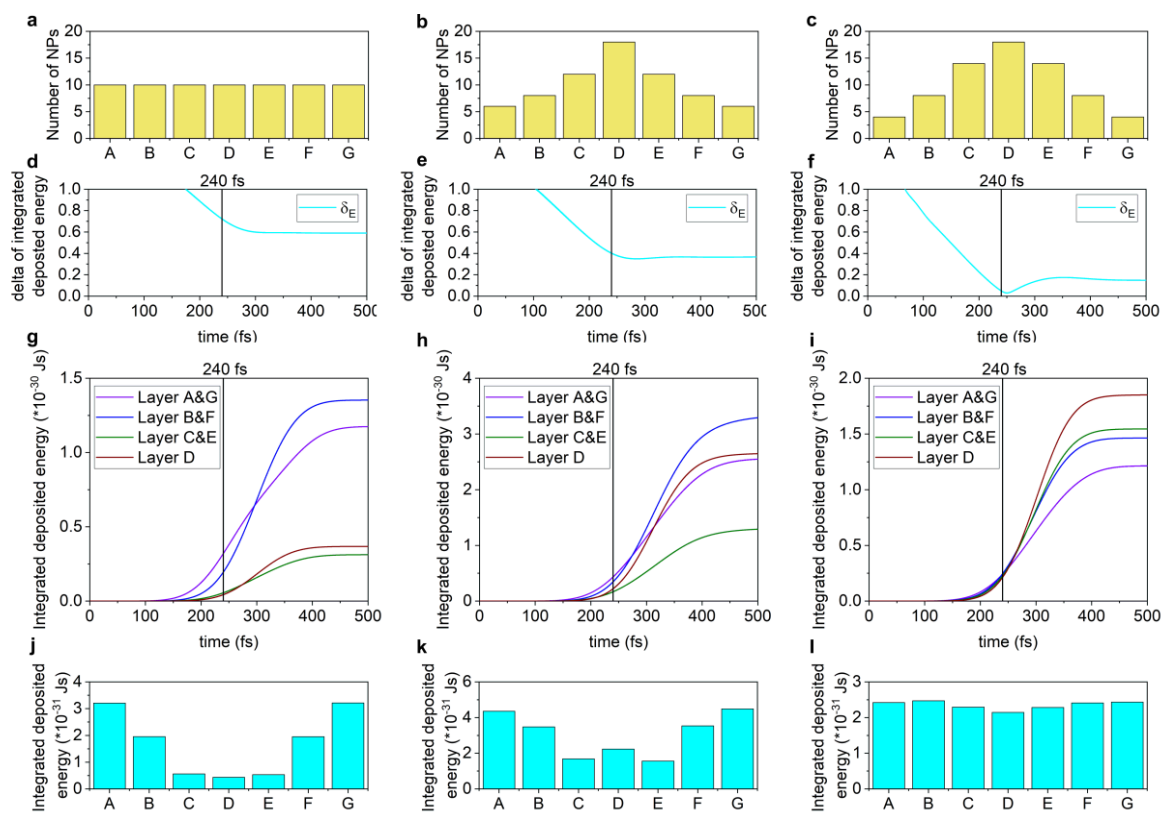

**Figure S1.** Time-dependent deposited energy in passive targets. The (a-c) nanoprism number density distribution along the target. The time-evolution of the (d-f) standard deviation of the deposited energy and the (g-i) integrated deposited energy. The distribution of the integrated deposited energy in different layers at 240 fs. (a, d, g, j) uniform, (b, e, h, k) Gaussian and (c, f, i, l) adjusted distributions.

The uniform distribution shows the largest minimal standard deviation in the deposited energy with the largest delay (Supplementary Table S1, Fig. S1a, d, g). The Gaussian distribution is better due to the intermediate minimal standard deviation taken with intermediate delay in the deposited energy (Supplementary Table S1, Fig. S1b, e, h). The advantage of the adjusted distribution is the smallest minimal standard deviation in the deposited energy, taken with the smallest delay (Supplementary Table S1, Fig. S1. c, f, i).

*Evaluation at the time-instant of pulse-overlap*

The integrated deposited energy is the smallest in uniform nanoresonator distribution, moreover the standard deviation at 240 fs is the largest (Supplementary Table S1, Fig. S1. g, j). In case of the Gaussian distribution the deposited energy is the largest and the standard deviation is intermediate (Supplementary Table S1, Fig S1. h, k). The adjusted distribution shows intermediate deposited energy, and the smallest standard deviation (Supplementary Table S1, Fig. S1. i, l)

Based on the FOM of the deposited energy, the uniform distribution ($2.35 \times 10^{-31}$ Js) is the weakest, the Gaussian distribution is intermediate ($7.60 \times 10^{-31}$ Js), while the adjusted distribution ($4.79 \times 10^{-30}$ Js) is the most advantageous (Supplementary Table S1).

In summary, the uniform distribution is the weakest in the average value, standard deviation and the FOM of the deposited energy. The Gaussian distribution is intermediate in the standard deviation and in the FOM, and the most advantageous in the average deposited energy. The adjusted distribution is the most advantageous on the average, due to its smallest standard deviation and largest FOM, though the average value is intermediate.

*Ranking of the passive targets*

If every inspected quantity is equally considered in the ranking, namely counting all quantities, then the distributions are not comparable. The weakest / compromised intermediate / the most preferable is the uniform / Gaussian / adjusted distribution, as it shows 5 – 0 – 0 / 0 – 4 – 1 / 0 – 1 – 4 quantities; in which the specific system is the weakest – intermediate – the most preferable. Based on the FOM, the distribution ranking shows uniform – Gaussian – adjusted order, so the most advantageous is the adjusted distribution, in accordance with the intuitive expectations.

**Active targets**

*Dynamics of standard deviation*

The uniform distribution in active target shows the largest minimal standard deviation in the deposited energy with intermediate delay (Supplementary Table S1, Fig. S2a, d, g). Compared to the uniform distribution in passive target, the minimal standard deviation is increased, while the delay is decreased. The Gaussian distribution in active target is advantageous due to the intermediate minimal standard deviation in the deposited energy, but it has the largest delay (Supplementary Table S1, Fig. S1b, e, h). Compared to the Gaussian distribution in passive target, both the minimal standard deviation and its delay is increased. The advantage of the adjusted distribution is that the smallest minimal standard deviation in the deposited energy is achieved with the smallest delay in the active targets, as a result the temporal characteristics are the most advantageous (Supplementary Table S1, Fig. S2. c, f, i). Compared to the adjusted distribution in the passive target, the minimal standard deviation is increased in the deposited energy, while the delay of it is decreased.

*Evaluation at the time-instant of pulse-overlap*

The integrated deposited energy is the smallest in uniform nanoresonator distribution, moreover the standard deviation is the largest (Supplementary Table S1, Fig. S2. g, j). In case of the Gaussian distribution the deposited energy takes on the largest value, while its standard deviation is intermediate (Supplementary Table S1, Fig. S2. h, k). The adjusted distribution resulted in intermediate deposited energy and the smallest standard deviation (Supplementary Table S1, Fig. S2. i, l) In case of uniform and Gaussian distributions in active target, the deposited energy is smaller, while the standard deviation is larger compared to their counterparts in passive target.

In the adjusted distribution though the deposited energy is decreased, its standard deviation is also smaller, which indicates that the dye doping is advantageous in achieving more uniform deposited energy distribution along the target.

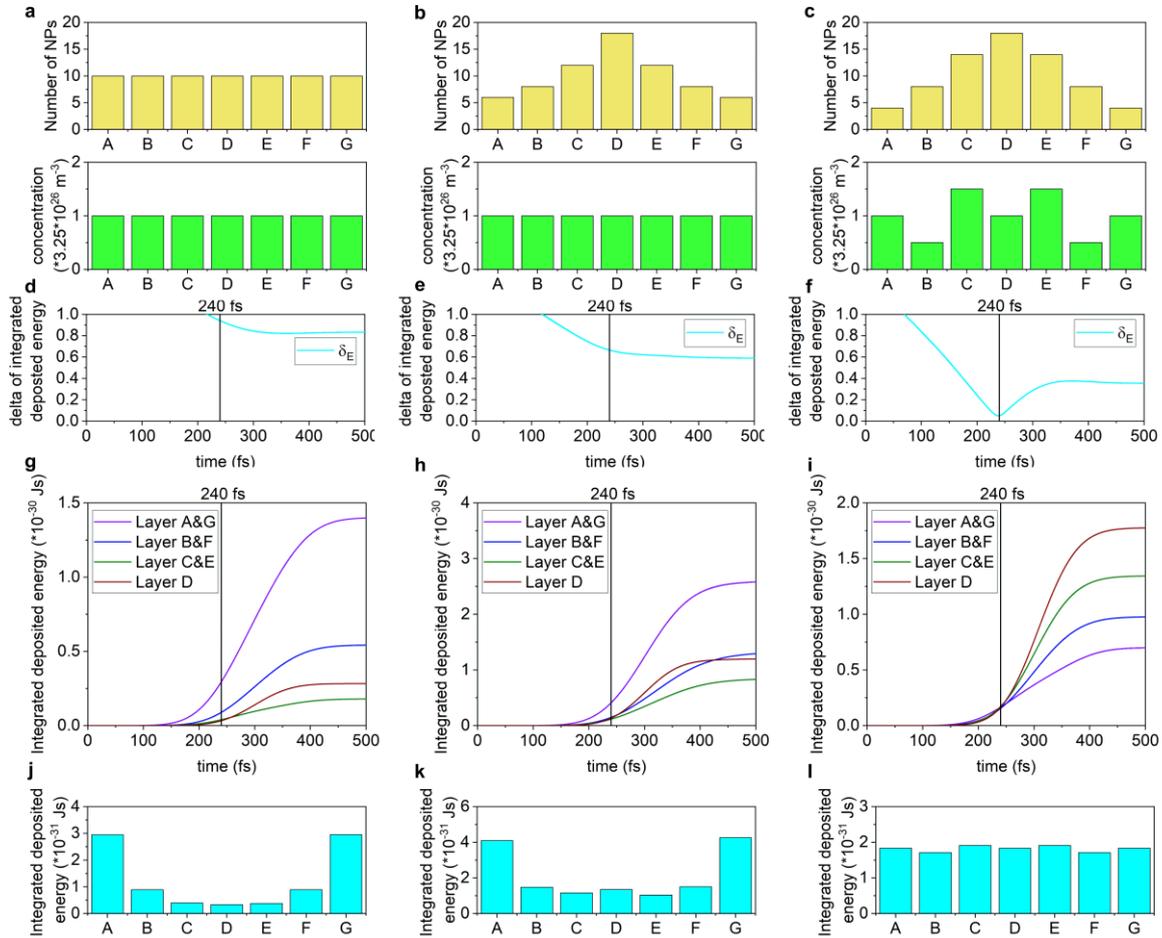

**Figure S2.** The time-dependent deposited energy in active targets. The (a-c) nanoprism number density and dye molecule concentration distribution along the target. The time-evolution of the (d-f) standard deviation of the deposited energy and the (g-i) integrated deposited energy. The distribution of the integrated deposited energy in different layers at 240 fs. (a, d, g, j) uniform, (b, e, h, k) Gaussian and (c, f, i, l) adjusted distributions.

Based on the FOM of the deposited energy the uniform distribution is the weakest (1.33×10$^{-31}$ Js), the Gaussian distribution is intermediate (3.20×10$^{-31}$ Js), and the adjusted distribution is the most advantageous (3.78×10$^{-30}$ Js) (Supplementary Table S1). The FOM of the energy is smaller in every inspected distribution compared to their counterparts in passive target. Similarly to the passive targets, the adjusted distribution is the most advantageous also in case of the active targets.

Uniform nanoresonator distribution is the least advantageous in the active target because of the smallest deposited energy, FOM of the deposited energy, and the largest standard deviations of the deposited energy.

The Gaussian distribution in the active target allows for the largest deposited energy, while it is intermediate in its standard deviations and FOM.

The unambiguous advantage of the adjusted distribution in the active target is indicated by the largest FOM and smallest standard deviation of the deposited energy, though it is intermediate in the average deposited energy.

Based on the deposited energy, the inspected distributions in passive targets outperform those in the active targets in the average value, the standard deviation and also in the FOM, except in the adjusted distribution in the active target, which allows for standard deviation smaller than that achievable in counterpart distribution in a passive target.

*Ranking of the active targets*

Similarly to the passive targets, the ranking is not balanced, when every inspected quantity is equally considered, the only difference is that the adjusted distribution becomes more advantageous, than the Gaussian distribution. Namely, the weakest / compromised intermediate / the most advantageous is the uniform / Gaussian / adjusted, as it shows 4 – 1 – 0 / 1 – 3 – 1 / 0 – 1 – 4 quantities; in which the specific system is the weakest – intermediate – the most preferable. Based on the FOM, the distribution ranking shows uniform – Gaussian – adjusted order, so the most advantageous is the adjusted distribution, in accordance with the intuitive expectations.

| Targets seeded with gold nanoprisms | | | | | | |
|---|---|---|---|---|---|---|
|  | Au-P | Au-SP-P | Au J-P | Au-A | Au-SP-A | Au-J-A |
| $\delta_{min\_PL}$ | 0.594 | 0.353 | 0.029 | 0.831 | 0.601 | 0.048 |
| $t_{min\_PL}$ (fs) | 499 | 288 | 250 | 361 | 500 | 240 |
| $\Delta t_{min\_PL}$ (fs) | 259 | 48 | 10 | 121 | 260 | 0 |
| $PL_{240fs}$ (J) | $2.59*10^{-17}$ | $6.82*10^{-18}$ | $3.61*10^{-17}$ | $1.89*10^{-17}$ | $4.78*10^{-18}$ | $2.77*10^{-17}$ |
| $\delta_{PL\_240fs}$ | 0.724 | 0.407 | 0.051 | 0.946 | 0.674 | 0.048 |
| $FOM_{PL}$ (J) | $3.58*10^{-17}$ | $1.68*10^{-17}$ | $7.14*10^{-16}$ | $1.99*10^{-17}$ | $7.02*10^{-18}$ | $5.76*10^{-16}$ |
| $\delta_{min\_NFE}$ | 0.165 | 0.170 | 0.071 | 0.136 | 0.156 | 0.015 |
| $t_{min\_NFE}$ (fs) | 290 | 249 | 223 | 277 | 272 | 430 |
| $\Delta t_{min\_NFE}$ (fs) | 50 | 9 | 17 | 37 | 32 | 190 |
| $NFE_{240fs}$ | 3.94 | 6.55 | 8.61 | 3.23 | 6.83 | 6.74 |
| $\delta_{NFE\_240fs}$ | 0.461 | 0.358 | 0.330 | 0.360 | 0.51 | 0.093 |
| $FOM_{NFE}$ | 8.54 | 18.29 | 26.08 | 8.97 | 13.63 | 72.35 |
| $\delta_{min\_E}$ | 0.590 | 0.349 | 0.029 | 0.821 | 0.588 | 0.048 |
| $t_{min\_E}$ (fs) | 500 | 285 | 250 | 361 | 500 | 240 |
| $\Delta t_{min\_E}$ (fs) | 260 | 45 | 10 | 121 | 260 | 0 |
| $E_{240fs}$ (Js) | $1.69*10^{-31}$ | $3.05*10^{-31}$ | $2.35*10^{-31}$ | $1.26*10^{-31}$ | $2.12*10^{-31}$ | $1.80*10^{-31}$ |
| $\delta_{E\_240fs}$ | 0.721 | 0.401 | 0.049 | 0.940 | 0.664 | 0.048 |
| $FOM_E$ (Js) | $2.35*10^{-31}$ | $7.60*10^{-31}$ | $4.79*10^{-30}$ | $1.33*10^{-31}$ | $3.20*10^{-31}$ | $3.78*10^{-30}$ |

**Table S1.** The minimal standard deviation ($\delta_{min\_PL}$, $\delta_{min\_NFE}$, $\delta_{min\_E}$), time instant ($t_{min\_PL}$, $t_{min\_NFE}$, $t_{min\_E}$) and delay ($\Delta t_{min\_PL}$, $\Delta t_{min\_NFE}$, $\Delta t_{min\_E}$) of it, the average value along the target ($PL_{240fs}$, $NFE_{240fs}$, $E_{240fs}$) and its standard deviation ($\delta_{PL\_240fs}$, $\delta_{NFE\_240fs}$, $\delta_{E\_240fs}$) at 240 fs and the FOM of the power-loss ($FOM_{PL}$), NFE ($FOM_{NFE}$) and deposited energy ($FOM_E$) in case of passive / active uniform (Au-P / Au-A), single-peaked Gaussian (Au-SP-P / Au-SP-A) and adjusted (Au-J-P / Au-J-A) nanoprism distribution. Color legend: black / grey is the weakest, blue / green is intermediate and red / orange is the most advantageous distribution in passive / active targets. Green background indicates where the active targets are better than passive ones.

| $\delta_{min\_PL}$ | $\delta_{min\_NFE}$ | $\delta_{min\_E}$ | $\Delta t_{min\_PL}$ | $\Delta t_{min\_NFE}$ | $\Delta t_{min\_E}$ |
|---|---|---|---|---|---|
| Au-J-P | Au-J-A | Au-J-P | Au-J-A | Au-SP-P | Au-J-A |
| Au-J-A | Au-J-P | Au-J-A | Au-J-P | Au-J-P | Au-J-P |
| Au-SP-P | Au-A | Au-SP-P | Au-SP-P | Au-SP-A | Au-SP-P |
| Au-P | Au-SP-A | Au-SP-A | Au-A | Au-A | Au-A |
| Au-SP-A | Au-P | Au-P | Au-P | Au-P | Au-P |
| Au-A | Au-SP-P | Au-A | Au-SP-A | Au-J-A | Au-SP-A |

**Table S2.** The ranking of the inspected targets in the minimal standard deviation and in its delay for the power-loss, *NFE* and the deposited energy in case of passive / active uniform (Au-P / Au-A), single-peaked Gaussian (Au-SP-P / Au-SP-A) and adjusted (Au-J-P / Au-J-A) gold nanoprism distribution. Color legend: black / grey is the weakest, blue / green is intermediate and red / orange is the most advantageous distribution in passive / active targets. Green background indicates where the active targets are better than passive ones.

| $PL_{240fs}$ | $NFE_{240fs}$ | $E_{240fs}$ | $\delta_{PL\_240fs}$ | $\delta_{NFE\_240fs}$ | $\delta_{E\_240fs}$ | $FOM_{PL}$ | $FOM_{NFE}$ | $FOM_E$ |
|---|---|---|---|---|---|---|---|---|
| Au-J-P | Au-J-P | Au-SP-P | Au-J-A | Au-J-A | Au-J-A | Au-J-P | Au-J-A | Au-J-P |
| Au-J-A | Au-SP-A | Au-J-P | Au-J-P | Au-J-P | Au-J-P | Au-J-A | Au-J-P | Au-J-A |
| Au-P | Au-J-A | Au-SP-A | Au-SP-P | Au-SP-P | Au-SP-P | Au-P | Au-SP-P | Au-SP-P |
| Au-A | Au-SP-P | Au-J-A | Au-SP-A | Au-A | Au-SP-A | Au-A | Au-SP-A | Au-SP-A |
| Au-SP-P | Au-P | Au-P | Au-P | Au-P | Au-P | Au-SP-P | Au-A | Au-P |
| Au-SP-A | Au-A | Au-A | Au-A | Au-SP-A | Au-A | Au-SP-A | Au-P | Au-A |

**Table S3.** The ranking of the inspected targets in the average value along the target and its standard deviation at 240 fs and the FOM of the power-loss, *NFE* and deposited energy in case of passive / active uniform (Au-P / Au-A), single-peaked Gaussian (Au-SP-P / Au-SP-A) and adjusted (Au-J-P / Au-J-A) gold nanoprism distribution. Color legend: black / grey is the weakest, blue / green is intermediate and red / orange is the most advantageous distribution in passive / active targets. Green background indicates where the active targets are better than passive ones.